# Mining Mathematical Documents for Question Answering via Unsupervised Formula Labeling

Philipp Scharpf[1], Moritz Schubotz[2], and Bela Gipp[3]

[1]University of Konstanz, Germany ({first.last}@uni-konstanz.de)
[2]FIZ-Karlsruhe, Germany ({first.last}@fiz-karlsruhe.de)
[3]University of Göttingen, Germany ({last}@cs.uni-goettingen.de)

April 6, 2022

**Abstract**

The increasing number of questions on Question Answering (QA) platforms like Math Stack Exchange (MSE) signifies a growing information need to answer math-related questions. However, there is currently very little research on approaches for an open data QA system that retrieves mathematical formulae using their concept names or querying formula identifier relationships from knowledge graphs. In this paper, we aim to bridge the gap by presenting data mining methods and benchmark results to employ Mathematical Entity Linking (MathEL) and Unsupervised Formula Labeling (UFL) for semantic formula search and mathematical question answering (MathQA) on the arXiv preprint repository, Wikipedia, and Wikidata. The new methods extend our previously introduced system , which is part of the Wikimedia ecosystem of free knowledge. Based on different types of information needs, we evaluate our system in 15 information need modes, assessing over 7,000 query results. Furthermore, we compare its performance to a commercial knowledge-base and calculation-engine (Wolfram Alpha) and search-engine (Google). The open source system is hosted by Wikimedia at `https://mathqa.wmflabs.org`. A demovideo is available at `purl.org/mathqa`.

## 1 Introduction

A large part of mathematical search queries are formulated as well-formed questions [20]. Factoid question answering systems allow the user to pose questions in natural language to provide quick and concise answers. In contrast, search engines typically display ranked lists of web pages or documents [10]. Semantic search engines aim to 'understand' the meaning and intent of a user's query instead of just retrieving literal or fuzzy matches of the input words [4].

Figure 1: MathQA semantic search example relationship question with identifier name and value retrieval and calculation. Demovideo at `purl.org/mathqa`.

In this paper, we continue our research on semantic formula search and factoid mathematical question answering using our open-source *MathQA* system[1], which is hosted by Wikimedia at `https://mathqa.wmflabs.org`. The prominent novelty of our contribution is the open source publication of a comprehensive and detailed benchmark for semantic formula search and mathematical question answering on open data sources. We extend our former work [39] by a comprehensive system evaluation in 15 different information need modes [5], out of which only one was previously available. Furthermore, we add a comparison to two state-of-the-art commercial competitors (Wolfram Alpha and Google). The system can answer mathematical questions in English and Hindi language, taking formula concept names and identifier relationships as intents. Identifiers are formula variables with no fixed value[2]. Besides numbers and operators, they are one of several formula constituent types. For example, the physics formula $F = m \cdot a$ contains the identifiers $F$, $m$, and $a$. The system (Figure 1) presents the succinct formula answer for an identifier relationship question along with names for the formula identifiers so that the user can understand their meaning. Besides, values for constants are retrieved from the semantic knowledge-base Wikidata, if available. Using these and additional user input for the variables, MathQA also allows for calculations.

Wikidata was launched in 2012 to support Wikipedia by providing language-independent items containing factual information that is framed as claims [44]. The claims consist of item-property relationship statements, which should be supported by sources and can be read, accepted, declined, or edited by humans and bots. Up to now, Wikidata contains around 5,000 statements[3] that link an item concept name to a mathematical formula [30]. Our MathQA system exploits this information along with semantic indices, which we created from the NTCIR Wikipedia and arXiv datasets [2] (unsupervised retrieval without annotation). These datasets contain a selection of documents and articles to be used as benchmarks for Mathematical Information Retrieval (MathIR) tasks. Our unsupervised approach differs from supervised math problem solving experiments, such as [19, 3] by mining linked open data (Wikidata) and open access corporae (NTCIR arXiv and Wikipedia). Moreover, it can not be compared to traditional formula search engines that search formula names and resources, taking the formula string as input. Here we focus on the opposite way, performing a 'semantic formula search' by retrieving formula strings from names.

---

[1] A demovideo is available at `purl.org/mathqa`
[2] `https://www.w3.org/TR/MathML3/chapter4.html#contm.ci`
[3] Run `https://w.wiki/z8p` to get the current number.

## 2 Related Work

In the following, we describe the state of the art in Factoid question answering and mathematical QA systems. MathQA is a factoid QA system, and since we concentrate our evaluations on the physics domain[4], we review related systems.

### 2.1 Factoid Question Answering Systems

Factoid question answering systems, providing fast and succinct answers [10], typically employ open semantic knowledge bases such as Freebase [8] or Wikidata [44] for answer retrieval. They are evaluated on datasets that contain labeled question-answer pairs, which refer to resources in the open databases [6]. Besides the challenges of knowledge base population [16], it is costly to generate large benchmark datasets.

**Datasets** Since the start of the QA Track [5] at the 'Text REtrieval Conference' (TREC-8) in 1998, there have been efforts to build QA systems and datasets [24]. Berant et al. introduce the 'WebQuestions dataset' for benchmarking QA engines that work on structured knowledge bases [6]. The 'Stanford Question Answering Dataset' (SQuAD) [28] contains 100,000 questions posed by crowdworkers on a set of Wikipedia articles. Bordes et al. introduce the 'SimpleQuestions dataset' containing 108,442 simple questions over Freebase triples (subject, predicate, object) [9]. The 'WikiQuestions dataset' contains 4,390,597 questions and corresponding answer entities, generated by rephrasing Wikipedia sentences as questions using a Wikipedia dump with Freebase entity mentions [27]. Applying a novel neural network architecture on Freebase to transduce facts into natural language questions, Serban et al. are able to generate 30 Million questions for the '30M Factoid Question-Answer corpus' [40].

**Systems** Knowledge Graph based Question Answering (KG-QA) aims to answer natural language questions retrieving facts from a knowledge graph [14]. Recent approaches employ neural networks for question generation [40] or answer retrieval. Besides recurrent architectures, also long-term memory networks [9] or convolutional neural networks [45] are used for large-scale simple question answering. In 2018, Tanon et al. introduce 'Platypus' as a multilingual question answering platform for Wikidata [43]. The system can answer complex queries in several languages, using hybrid grammatical and template based techniques [39]. The 'MathQA' system is based on Platypus with a focus on the mathematics domain.

---

[4]Note that in our last publication, we already presented a general evaluation on random math domains. Here, we focus on physics, having domain experts for the assessment.

[5]https://trec.nist.gov/data/qamain.html

## 2.2 Mathematical and Physics Question Answering

Question answering in the domain of mathematics was first implemented by Smith in 1974. Investigating the understanding of natural language by computers, a system to answer elementary mathematics questions using 'unrestricted natural language input' was implemented [41]. Unfortunately, until the last decade, there was little interest and progress in the subject of MathQA. In 2012, Nguyen et al. introduced a math-aware search engine for a math question answering system [23]. Their system can handle both textual keywords and mathematical expressions. They use a Finite State Machine model to encode the semantics of mathematical expressions and an online learning binary classifier for the ranking. The approach was benchmarked against three classical information retrieval (IR) strategies on math documents crawled from Math Overflow, claiming other methods by more than 9%. In 2017, Bhattacharya et al. published a survey of question answering for math and science problems [7]. They review past and present efforts to make computers smart enough to pass math and science tests. They conclude that 'the smartest AI could not pass high school.' In the 'SemEval 2019 task' on math question answering, Hopkins et al. derive a question set from practice exams [13]. Using 2778 training questions, the top system could answer 45% of the 1082 test questions correctly, significantly better than the random guessing baseline found at 17%. In 2018, Gunawan et al. introduced an Indonesian question answering system for solving arithmetic word problems using pattern matching, which was integrated into a physical humanoid robot [12]. Characterizing searches for mathematical concepts, Mansouri et al. investigate search engine queries to find that well-formed questions were surprisingly common [20]. This was one motivation for the 'ARQMath Lab' at CLEF 2020. In two tasks, the goal was to find answers to new mathematical questions posted on a community question answering site (Math Stack Exchange) by referring to old QA threads, containing both text and formulae [21, 46, 34]. As the results indicated that approaches to the challenging tasks still need to be elaborated further, the ARQMath Lab is planned to be continued in the coming years.

Compared to general mathematical question answering, even less research is done on the physics domain. Pineau [26] and Abdi et al. [1] discuss and present first approaches to answer questions on physics. Pineau claims that equations encapsulate a crucial part of the knowledge in physics [26]. Since equations can be connected via their natural language meaning, the need for a semantic search on physics is implied. Furthermore, a cross-disciplinary search would allow researchers to find solutions to their problems or equations in other fields. In 2018, Abdi et al. introduced an ontology-based question answering system in the physics domain (QAPD) [1]. In the first step, an ontology is populated using information from a textbook, lecture notes, and course materials. Secondly, ten human experts are asked to generate entity-relationship questions on ontology knowledge. These include identifier definition and unit queries, such as 'how to calculate resistance?' and 'what is the unit of resistance?'. The system is evaluated on 3750 queries, achieving an F-measure of 76%. Since units of measurement are an essential part of physical calculations, there have been efforts to automatically infer them from articles (Wikipedia) supported by knowledge-base groundings (Wikidata) [36].

**Mathematical Entity Linking** Kristianto et al. propose methods to link mathematical expression in scientific documents to Wikipedia articles using their surrounding text [18, 17]. Their learning-based approach achieves a precision of 83%, compared with a 6.22 baseline of a traditional MathIR method. A balanced combination of mathematical and textual elements is required for the linking performance to be reliable.

Besides linking to Wikipedia, Schubotz et al. [38, 32] describe linking mathematical formula content to Wikidata, both in MathML and LaTeXmarkup. To extend classical citations by mathematical, they call for a 'Formula Concept Discovery (FCD) and Formula Concept Recognition (FCR) challenge' to elaborate automated MathEL. Their FCD approach yields a recall of 68% and precision of 68% for retrieving equivalent representations of frequent formulae. In 72% of the cases, a formula name could be extracted from the surrounding text on the NTCIR arXiv dataset [2].

**Applications** Mathematical Entity Linking - being less popular than its natural language correspondent - has so far been employed in mathematical question answering systems, such as 'MathQA' using structured Wikidata items [39] and proposed for semi-structured question posts from Math Stack Exchange (MSE) at the CLEF ARQMath Lab [34]. Moreover, it is expected that MathEL will enhance mathematical subject classification [35, 37].

## 3 Methods

Having reviewed the related work literature (see Section 2), we identify the ability of a mathematical QA system to answer identifier[6] relationship questions, e.g., 'what is the relationship between mass and energy', as research gap. With a research objective to investigate the feasibility of identifier relationship question answering, our experiments were driven by the following research questions:

1. What is the quality of a translation of identifier symbols to names and vice versa?

2. How well does a semantic formula search using identifier names or symbols as query perform?

3. How well does a semantic formula search using the formula concept name as query perform?

---

[6] For a definition of an identifier, see the introduction.

4. How well do different index sources created from different datasets perform in comparison?

5. How well does the semantic search perform compared to a commercial knowledge-base and search engine?

In the following, we describe the methods we employed to answer the research questions.

### 3.1 MathQA Workflow

Our mathematical question answering system workflow consists of the following steps:

1. question parsing and classification,
2. index or knowledge-graph query,
3. entity linking or relationship extraction,
4. answer (candidate) retrieval and presentation,
5. formula parsing, and
6. result calculation.

In some cases, a QA system involves an additional question domain classification step. Since we concentrate on mathematical question answering and MathQA does not distinguish subject classes, we skip this step. The implementation of these steps in our system will be described in Section 4. In this section, we discuss the high-level concepts that are involved.

### 3.2 Question Parsing and Classification

MathQA is designed to answer the following questions:

1. What is the formula for [formula name]?
2. What is the [property] of [geometric object]?
3. What is the relationship between [identifier name 1] and [identifier name 2] and ...?
4. What is the relationship between [identifier symbol 1] and [identifier symbol 2] and ...?

Question (1) is a general math type question yielding formula concepts, whereas question (2) is a geometry type question, e.g., 'what is the area of a circle?'. Given that there are different question types, the first step in question parsing is to distinguish them. Question types (3) and (4) can be easily recognized by the keywords 'relationship' or 'relation.' To differentiate (1) and (2), we need to transform the question into a tree of triples (subject, predicate, object). For type (1), the predicate is 'formula,' and the object needs to be retrieved: (subject, formula,?), e.g., (velocity, formula,?). For type (2), both predicate and subject are variable, yielding: (geometrical object, property,?), e.g., (sphere, volume,?). We retrieved a list of geometry properties (volume, area, radius, etc.) from Wikipedia. The system checks if the triple predicate is in this list to classify the question as geometry type (2).

### 3.3 Mathematical Entity Linking

To answer the question types (1)-(4), entity (identifier or formula) concept names need to be linked to symbols or strings. Using Wikidata, an additional entity identification number (QID) is available. For example, the identifier symbol 'E' can be linked to the Wikidata item with the name 'energy' and QID 'Q11379' if it occurs in the formula '$E = mc^2$', which can be assigned to the concept name 'mass-energy equivalence' (Q35875). In our evaluation, we compared three different sources for Math Entity Linking: indices created using document / article selections taken from 1) the preprint repository 'arXiv'[7] or 2) Wikipedia, or knowledge-graph content retrieved from 3) Wikidata using SPARQL queries[8]. We will discuss the index creation of 1) and 2) in Section 3.5 and querying Wikidata in Section 4, where we describe the Formula Retrieval Module of MathQA.

### 3.4 Benchmarking

To evaluate MathIR methods and systems, such as MathQA, we need benchmark samples and datasets. We will now introduce the sources that are relevant to our studies. As stated in the introduction, we exploit the NTCIR 11/12 arXiv, and Wikipedia dataset [2] to create our semantic indices for the Formula Retrieval Module. The dataset is available at http://ntcir-math.nii.ac.jp/data. It consists of 105,120 document sections taken arXiv papers, in total containing over 60 million mathematical formulae in MathML markup. The Wikipedia articles were converted from Wikitext to HTML.

Our evaluation sample consists of formula concepts, which were annotated using the *AnnoMathTeX*[9] formula and identifier annotation recommender system [33, 29]. The formulae were taken from an already existing benchmark selection of 25 Wikipedia articles from physics (classical mechanics). The gold-standard is persisted in the *MathMLben*[10] repository [38].

---

[7] https://arxiv.org
[8] https://www.w3.org/TR/rdf-sparql-query
[9] https://annomathtex.wmflabs.org
[10] https://mathmlben.wmflabs.org

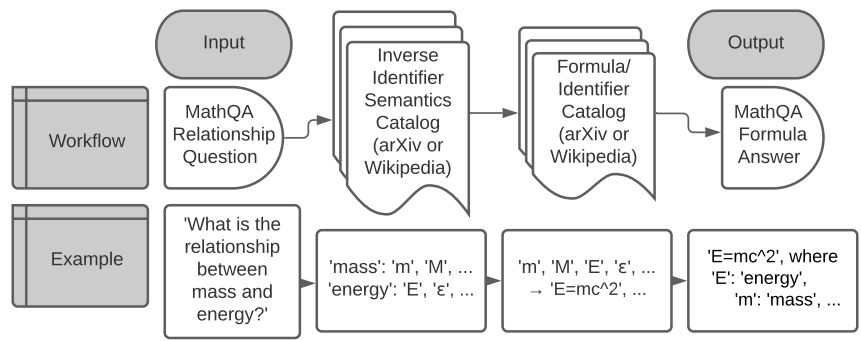

Figure 2: Workflow of MathQA answering a relationship question using a semantic identifier index (arXiv or Wikipedia) and formula catalog compiled from the NTCIR 11/12 datasets (arXiv or Wikipedia).

## 3.5 Semantic Formula and Identifier Indexing

There are two ways to create a semantic formula or identifier index vocabulary: 1) supervised labeling using Mathematical Entity Linking with systems such as AnnoMathTeX or 2) unsupervised extraction from corpora, such as the NTCIR arXiv and Wikipedia dataset. In this project, we use both methods in comparison. Specifically, we test an unsupervised formula and identifier index on a supervised benchmark sample. The index was created using Unsupervised Formula Labeling, which should not be confused with Latent Semantic Indexing (LSI) [11], which indexes documents instead of formulae or identifiers.

We created the following index catalogs:

1. identifier-semantics catalog (symbol to name)[11]
2. semantics-identifier catalog (name to symbol)[12]
3. formula concept catalog (name to string)[13]

We extracted (1) and (3) from the corpus and then inverted (1) to obtain (2). For each formula or identifier in the corpora, we attributed words in the surrounding text and ranked them by the frequency of their occurrence (see the scores sorted by subject class[14]).

Figure 2 shows how the formula and identifier catalogs are employed in the semantic search of MathQA to answer an identifier relationship question.

---

[11] For example, 'm' to 'mass'.
[12] For example, 'force' to 'F'.
[13] For example, 'momentum' to 'p = m v'.
[14] https://en.wikipedia.org/wiki/User:Physikerwelt?oldid=738857609

**Identifier Catalogs**  We extracted the identifier from MathML `<mi>` tags using a word window of ± 500 characters, which was chosen such that the average MathML string length is overcome and words outside the formula environment are reached. We neglected identifier indices. The terms were lowered and cleaned from punctuation or other symbols. Stopwords were excluded. The Wikipedia identifier semantics catalog contains a total of 1670 entries, whereas the arXiv catalog contains 94833. The difference (factor 57) is reasonable considering the different corpus sizes in terms of documents (factor 53).

**Formula Catalogs**  The number of formulae in the catalogs are 30776 for Wikipedia and 118120 for arXiv. As we were evaluating on a physics sample (see Section 5), we confined the arXiv index to 10 physics subject classes: 'astro-ph', 'cond-mat', 'gr-qc', 'hep-lat', 'hep-ph', 'hep-th', 'math-ph', 'nlin', 'quant-ph', and 'physics'. The total number of formulae in the catalog is 134217 for Wikipedia and 3450770 for arXiv (here the increase is only a factor 26). We created single-word indices from the surrounding text of the formulae. For multiple word concept queries, such as 'angular acceleration', we joined the results in a union.

## 4  Implementation

While in Section 3, we explained the data mining for MathQA, in this section we will describe the system with its constituent modules. According to the steps presented in Section 3.1, we developed five MathQA modules. The Question Parsing Module (step 1) transforms questions into a triple representation to classify the type of intent (general, geometry, or relationship). The Formula Retrieval Module (steps 2 and 3) queries an index (arXiv, Wikipedia) or knowledge-graph (Wikidata). The Formula Answer Module (step 4) identifies the candidates and presents the top formula result to the user, including identifier names and values if available. The Formula Parsing and Calculation Module (steps 5 and 6) split the formula into its constituents to allow for calculation using the user's input values for variables and retrieved values for constants if applicable. The MathQA web interface design is based on Ask Platypus[15]. The system relies on the web application framework Flask [16]. The programming languages Python, JavaScript, and HTML are used.

**Question Parsing Module**  Employing the Stanford CoreNLP[17] server and Natural Language Toolkit (NLTK)[18] library, the module produces (subject, predicate, object) triples from the free-text questions. The CoreNLP server runs in Java. It allows for the required dependency and constituency parsing in English. Currently, it supports five other languages: German, French, Spanish,

---

[15] https://askplatyp.us
[16] https://pypi.org/project/Flask
[17] https://stanfordnlp.github.io/CoreNLP
[18] https://www.nltk.org

```
# Find items with 'has part' (P527)
'energy' (Q11379) and 'mass' (Q11423)
SELECT ?item ?itemLabel ?formula ?parts ?partsLabel
WHERE {
    ?item wdt:P527 wd:Q11379.
    ?item wdt:P527 wd:Q11423.
    ?item wdt:P2534 ?formula.
    ?item wdt:P527 ?parts
SERVICE wikibase:label {
    bd:serviceParam wikibase:language "en".}}
```

Figure 3: SPARQL query to retrieve Wikidata formula items either using a python client or web interface at `https://query.wikidata.org`.

Arabic, and Chinese. The NLTK platform provides interfaces to corpora, lexical resources, and text processing methods. We use its parsing, tokenization, stemming, and stopword removal capabilities for our system.

**Formula Retrieval Module** We use Pywikibot[19] and SPARQL[20] libraries to retrieve formulae (format unrestricted) and their identifiers (amount unlimited) from Wikidata items. Pywikibot is an interface to the MediaWiki API, whereas the SPARQL client can perform SELECT and ASK queries against a SPARQL endpoint via HTTP. Figure 3 shows an example SPARQL query for an identifier relationship question to retrieve all items with specific items, e.g., 'energy' (Q11379), as 'has part' property (P527). A permanent speedlink to the query in a web interface is `https://w.wiki/39RQ`. The analogous query for items as 'calculated from' (P4934) can be accessed via `https://w.wiki/39RR`. We need to query both properties since, currently, they are both used by the community. Identifier symbols are commonly inserted via 'quantity symbol (LaTeX)' (P7973), 'quantity symbol (string)' (P416), or 'in defining formula' (P7235) property. You can find an example query at `https://w.wiki/$vr`. We employed the aforementioned queries in our evaluation tasks.

**Formula Parsing and Calculation Module** To provide calculations using the formula answer, the LaTeX string needs to be parsed. This is done via the python SimPy[21] module. Although the main purpose of the library is to provide process-based discrete-event simulations, it can also be used to generate abstract syntax trees from formulae and perform calculations. At the moment, only a single left-hand-side identifier can be calculated using input values for the right-hand side identifiers. In the future, we plan to use Computer Algebra Systems (CAS) to rearrange formulae such that each occurring identifier can be calculated.

---

[19] https://github.com/wikimedia/pywikibot
[20] https://pypi.org/project/sparql-client
[21] https://simpy.readthedocs.io/en/latest

Table 1: Excerpt of 10 from the 65 benchmark examples used for the evaluation of our MathQA system (modes 1-15). The goldstandard [38] is persisted at `https://mathmlben.wmflabs.org` (GoldID 310-375).

| GoldID | QID | Name | Formula |
|---|---|---|---|
| 310 | Q11376 | acceleration | $\mathbf{a} = \frac{\mathrm{d}\mathbf{v}}{\mathrm{d}t}$ |
| 311 | Q186300 | angular acceleration | $\boldsymbol{\alpha} = \frac{\mathrm{d}\boldsymbol{\omega}}{\mathrm{d}t}$ |
| 312 | Q834020 | angular frequency | $\omega = 2\pi f$ |
| 313 | Q161254 | angular momentum | $\mathbf{L} = \mathbf{r} \times \mathbf{p}$ |
| 314 | Q161635 | angular velocity | $\boldsymbol{\omega} = \frac{\mathrm{d}\varphi}{\mathrm{d}t}\mathbf{u}$ |
| 315 | Q2945123 | center of mass | $\sum_{i=1}^{n} m_i(\mathbf{r}_i - \mathbf{R}) = 0$ |
| 316 | Q2248131 | centripetal acceleration | $a_c = \frac{v^2}{r}$ |
| 317 | Q172881 | centripetal force | $\vec{F} = -\frac{mv^2 \hat{r}}{r}$ |
| 318 | Q843905 | circumference | $C = \pi \cdot d = 2\pi \cdot r$ |
| 319 | Q11382 | conservation of energy | $E_{\mathrm{tot1}} = E_{\mathrm{tot2}}$ |
| 320 | Q2305665 | conservation of momentum | $p_{\mathrm{tot1}} = p_{\mathrm{tot2}}$ |

## 5 Evaluation

To introduce benchmark results for mathematical question answering on open data, we evaluated our system on a formula set (Table 1), which was persisted on the MathMLben benchmark platform from [38]. Using graduated domain experts from physics, we assessed over 7,000 results in 15 different evaluation modes (as in [5]). Modes 1-6 (identifier semantics) are preparation steps for modes 7-12 (semantic search). All evaluation scripts and tables can be found in the respective mode folders in the MathQA repository at `https://github.com/ag-gipp/MathQA/tree/master/evaluation/semanticsearch`.

### 5.1 Evaluation Modes, Metrics, and Examples

In the following, we will describe the evaluation modes (input and output), metrics (accuracy, ranking), and examples (formula benchmark) we employed.

**Evaluation Modes** To evaluate the semantic search capabilities of our system, we assessed its performance to

- search identifier names by symbols,
- search identifier symbols by names,
- search formula strings by identifier symbols,
- search formula strings by identifier names,

Table 2: Evaluation mode 1-15 results for formula concept and identifier name or symbol queries using arXiv or Wikipedia indices or Wikidata SPAQRL queries. Only mode 15 (orange) was previously available. Modes 1-14 (blue) are contribution of this paper.

| Mode | Query | Top1 Acc. | mean(DCG) |
|------|-------|-----------|-----------|
| 1 | names to symbols, arXiv | 0.24 | 0.82 |
| 2 | names to symbols, Wikipedia | 0.31 | 0.92 |
| 3 | names to symbols, Wikdata | 0.12 | 0.20 |
| 4 | symbols to names, arXiv | 0.37 | 0.79 |
| 5 | symbols to names, Wikipedia | 0.12 | 0.54 |
| 6 | symbols to names, Wikdata | 0.22 | 0.20 |
| 7 | identifier names, arXiv | 0.06 | 0.23 |
| 8 | identifier names, Wikipedia | 0.03 | 0.24 |
| 9 | identifier names, Wikdata | 0.85 | 0.98 |
| 10 | identifier symbols, arXiv | 0.00 | 0.00 |
| 11 | identifier symbols, Wikipedia | 0.24 | 0.98 |
| 12 | identifier symbols, Wikdata | 0.48 | 0.46 |
| 13 | formula names, arXiv | 0.00 | 0.00 |
| 14 | formula names, Wikipedia | 0.17 | 0.98 |
| 15 | formula names, Wikdata | 0.52 | 1.03 |

- search formula strings by formula names.

Table 2 lists the resulting 15 different evaluation modes. The following evaluation sections will refer to the mode numbers. We will divide thematically into identifier names vs. symbols (modes 1-6), identifier relationship questions (modes 7-12), and formula concept name retrieval (modes 13-15).

For modes 1-15, we framed the evaluation as a ranking problem (calculating accuracy and ranking quality), aiming to determine the best source (arXiv, Wikipedia or Wikidata) to be used in our system. For mode 15, we additionally compared our system to commercial competitors (knowledge-base and calculation-engine Wolfram Alpha and search-engine Google).

**Evaluation Metrics** For each of our 15 different evaluation modes, we calculated the top1 accuracy and Discounted Cumulative Gain (DCG). In each mode and result, the system could score either 0 points (irrelevant), 1 point (relevant), or 2 points (exact match as in benchmark). We assessed the top10 results. Using this score, we could calculate the top1 accuracy as the number of results with score 1 or 2 divided by the total number of evaluations. The Discounted Cumulative Gain (DCG) ranking performance measure is calculated according

to [15] as

$$\text{DCG}_\text{p} = \sum_{i=1}^{p} \frac{rel_i}{\log_2(i+1)},$$

where $rel_i$ is the relevance (here 0, 1 or 2) at position i and p is the ranking scale cutoff (here position 10). In some cases, an Ideal Discounted Cumulative Gain (IDCG) can be set to calculate a normalized DCG (nDCG). In our case, we could not estimate an IDCG. One possibility would be to assign 2 points for each of the ten ranking positions. This would yield an IDCG of $\sum_{i=1}^{10} 2/log_2(i+1) = 9.09$. However, providing the exact benchmark match at each of the ten positions is very unlikely. Moreover, having the exact match (2 points) at position 1 and relevant hits (1 point) at each subsequent is not realistic either as ten identifier names or symbol synonyms often do not even exist. Each other possibility, e.g., 2 points first, followed by four times 1 point, and five times 0 points, is arbitrary. Therefore, we did not calculate an IDCG and nDCG.

**Evaluation Examples**  Our examples test set (Table 1) was created from a selection of 25 physics Wikipedia articles, for which formula and identifier entities were linked using a formula and identifier name annotation recommender system [33]. The formula selection is persisted on the benchmark platform MathMLben (https://mathmlben.wmflabs.org) ranging form GoldID 310 to 375. For each example formula, a GoldID represents the numbering. It also corresponds to the Wikidata QID of the concept item and its name. The constituting identifiers (e.g., E, m, and c for $E = mc^2$) are annotated and linked to Wikidata items using either the 'has part'(P527) or 'calculated from' (P4934) Wikidata properties. The formula and identifier names and symbols are used as query inputs, as described in the following subsections.

## 5.2  Formula Identifier Symbol and Name Relationships

Modes 1-6 prepare the evaluation of the semantic formula search (modes 7-12). It is assessed how accurate identifier names can be translated into symbols and vice versa. Example questions could be 'What are the symbols for energy?' or 'What are the meanings of the symbol E?'. To implement a semantic search on a formula database that is not semantically indexed, i.e., identifier names are not annotated, we need to translate the user's natural language query into symbolic language first (modes 1-3). Subsequently, after parsing, formulae can be found using their constituting identifier symbols. On the other hand, if we already have a semantic index, we can also query it using symbols after the translation (modes 4-6). In the following, we discuss the evaluation execution and results for the index sources arXiv, Wikipedia, and Wikidata, respectively.

**MathQA on arXiv and Wikipedia**  To evaluate the translation of identifier names to symbols and vice versa on the arXiv (modes 1 and 4) and Wikipedia (modes 2 and 5), we employed the respective semantic identifier indices, which were previously created from the NTCIR 11/12 arXiv dataset and Wikipedia (see Section 3.5). Each index is sorted by the ranking score (occurrence frequency of an identifier name-symbol relationship). For each example query formula from the benchmark (Table 1), all annotated identifier names and symbols are extracted. For each symbol or name, the top10 ranked results are then evaluated. For each result, we assess (score, rank) tuples to calculate the DCG ranking measure as described in Section 5.1. Table 3 shows example results for the arXiv (modes 1 and 4). With around 250 identifier name-symbol pairs, almost 500 query predictions (translation in both directions) had to be assessed. The evaluation is analogous for Wikipedia (modes 2 and 5) with an additional 500 evaluation table rows. The scripts and tables for prediction and scoring can be found in the respective mode folder of the MathQA repository.

Table 3: Example evaluation mode 1 and 4 results (first two GoldIDs) for identifier symbol-name relationship index that was created from the NTCIR arXiv 11/12 dataset [2]. Exact match to benchmark score 2 points, also relevant related results 1 point. The list (modes 1 and 4) is continued with 489 additional queries (total 498) and structurally identical to the one for Wikipedia (modes 2 and 5).

| GoldID | Query | Benchmark | Matches | (Score, Rank) | DCG |
|---|---|---|---|---|---|
| 310 | a | acceleration | - | - | 0 |
| 310 | v | velocity | velocity, vector, speed | (2,2), (1,5), (1,10) | 1.94 |
| 310 | t | duration | time | (1,1) | 1 |
| 310 | acceleration | a | a, g | (2,1), (1,4) | 1.69 |
| 310 | velocity | v | v, c, V, u | (2,1), (1,2), (1,4), (1,5) | 3.45 |
| 310 | duration | t | $\tau$ | (1,3) | 0.5 |
| 311 | $\alpha$ | angular acceleration | - | - | 0 |
| 311 | $\omega$ | angular velocity | frequency, oscillator, harmonic | (1,1), (1,8), (1,9) | 1.62 |
| 311 | t | duration | time | (1,1) | 1 |
| 312-375 | ... | ... | ... | ... | ... |

**MathQA on Wikidata** To evaluate the identifier name-symbol translation prediction of MathQA on Wikidata (modes 3 and 6), we employed SPARQL queries to retrieve the result candidates (see Section 4). First, a SPARQL query was compiled to find all items with 'quantity symbol (LaTeX)' (P7973) or 'quantity symbol (string)' (P416) or 'in defining formula' (P7235) in a union list. We had 84 results for P416, 1248 for P7973, and 599 for P7235, i.e., 1931 in total. Second, the items were used to create a semantic index for identifier names and symbols, respectively, as for the arXiv and Wikipedia. Since the Wikidata 'corpus' is smaller than the other source corporae, we only had results for 195 of the 500 queries. The missing ones were treated as scoring zeros. For mode 3 (names to symbols), the retrieved results scored a DCG of 0.25, yielding 0.20 taking into account the missing results. The top1 accuracy was 0.16 and 0.12, respectively. For mode 6 (symbols to names), the retrieved results scored a DCG of 0.34, yielding 0.20 taking into account the missing results. The top1 accuracy was 0.36 and 0.22, respectively. Comparing the two modes, it is apparent that the symbol-to-name conversion performs better than the name to symbol conversion. This contrasts the overall results for all sources (see next paragraph). Due to the small size of the Wikidata index, it may not be representative.

**Comparison** Table 2 shows the results of modes 1-6, comparing top1 accuracy and mean DCG. Comparing the mapping directions, we find that modes 1-3 ('names to symbols') perform better than modes 4-6 ('symbols to names') in terms of mean DCG (0.65 vs. 0.51), but slightly less in terms of top1 accuracy (0.22 vs. 0.23). This suggests the assumption that identifier symbols are more ambiguous than identifier names. For a given symbol, there are more potential names than there are symbols for a given name. Comparing the index sources, we find that modes 1 and 4 ('arXiv') perform better than modes 2 and 5 ('Wikipedia') and modes 3 and 6 ('Wikidata') both in terms of mean DCG (0.81 vs. 0.73 vs. 0.20) and top1 accuracy (0.30 vs. 0.21 vs. 0.17). The arXiv is most, Wikidata least efficient as a source to provide a semantic formula search by using identifier name or symbol relationships. Since the corpus size decreases from the arXiv to Wikipedia to Wikidata, this is an indication that corpus size helps to improve index quality (the larger, the better).

## 5.3 Formula Identifier Name or Symbol Relationship Questions

Modes 7-12 evaluate the semantic search of formulae by their constituting identifier names or symbols. Example questions could be 'What is the relationship between mass and energy?' or 'What is the relationship between the symbols m and E?'. For each mode and example, we evaluated the top10 ranked prediction of the different semantic indices for identifier and formulae (see Section 3.5). This yields a total of 6 x 66 = 396 queries to evaluate. As the formula indices are large, the semantic search provided more results, from which we only considered the top ten for each query.

**MathQA on arXiv and Wikipedia** Querying the NTCIR arXiv and Wikipedia formula indices for identifier relationships yielded a total 4225 formulae for the combinations of constituting identifiers (names or symbols). We also tested querying the formula index using multiple possibilities for each identifier name or symbol (e.g., symbols $E$ and $\epsilon$ for the name 'energy' - or names 'time' and 'duration for the symbol $t$). However, due to low accuracy, we discarded these modes. For each formula, the system also retrieved the name of the arXiv document (e.g., 'astro-ph0203007.tei') or Wikipedia article (e.g., 'Acceleration.html'). The number of results per example query (GoldID) varied between zero (e.g., for 'angular acceleration') to one (e.g., for 'electromagnetic force') to a maximum of 967 for 'momentum' (i.e., querying the relationship between momentum, mass, and velocity or p, m and v respectively). The fraction of the total retrieved formulae from the arXiv and Wikipedia indices is 4225 / 135997 = 3.1%.

**MathQA on Wikidata** Querying Wikidata for identifier relationships yielded a total 11285 formulae for the combinations of constituting identifiers (names or symbols). As for the identifier name and symbol translations (modes 1-6), we employed SPARQL queries. We retrieved identifier relationships via 'has part'(P527) or 'calculated from' (P4934) Wikidata item properties. Unfortunately, sometimes the right-hand side of a formula is not annotated, leading to fewer results for the name queries (mode 9). Fortunately, the symbol queries (mode 12) are not affected. They search for occurrences of the given symbols (resp. their combinations) in the mathml[22] formula strings of the Wikidata item's 'defining formula' property (P2534). The SPARQL queries yielded 10048 results for mode 9 but only 1237 for mode 12. Apparently, there are already a lot of items where the identifiers are annotated with their names. Due to the ambiguity of the identifier symbols, it is more favorable to query by names, and we suspect that users will also prefer this mode.

**Comparison** Table 2 shows the results of modes 7-12, comparing top1 accuracy and mean DCG. Comparing the query types, we find that querying by identifier names (modes 7-9) outperforms symbols (modes 10-12) in terms of top1 accuracy (0.31 vs. 0.24), while both types are equally good (0.48) in terms of mean DCG. A priori, we could not find a reason why one should outperform the other apart from the fact that symbols are more ambiguous than names since the vocabulary is much smaller. Comparing the index sources, we find that modes 9 and 12 ('Wikidata') perform better than modes 8 and 11 ('Wikipedia') and modes 7 and 10 ('arXiv') both in terms of mean DCG (0.72 vs. 0.61 vs. 0.12) and top1 accuracy (0.67 vs. 0.14 vs. 0.03). Here interestingly, a smaller index corpus size leads to more precision. This indicates that for the smaller corpora, the formula index is performing better than the identifier index.

### 5.4 Formula Concept Name Questions

Modes 13-15 evaluate the retrieval of formulae by their concept names (e.g., 'What is the formula for mass-energy equivalence?' yielding $E = mc^2$). While for the arXiv and Wikipedia (modes 13 and 14), we created a semantic index for formulae (in analogy to the identifier indices), for Wikidata (mode 15), the formulae are directly retrieved using a SPARQL query (see Section 4). Mode 15 is deployed in the live MathQA system, as it yielded the best results.

**MathQA on arXiv and Wikipedia** Querying the NTCIR arXiv and Wikipedia semantic formula index catalogs, we were confronted with the problem of retrieving very short formulae as top results (e.g., $t = 0$), which were mostly not relevant results. The reason is that formulae are ranked by the frequency of their occurrence (number of duplicates) in the corpus, and apparently, the short ones appear more often than the long ones. We tried to get more relevant results by inverting the ranking but could not improve the quality this way.

---

[22] https://www.w3.org/Math

Unfortunately, for the arXiv, the top1 accuracy and mean DCG is even zero, meaning that there were no relevant results (scoring 1 or 2) within the top10 hits. The Wikipedia index performed much better with a top1 accuracy of 0.17 (17% of the first hits were relevant) and mean DCG of 0.98, which is close to the performance when querying Wikidata (see the comparison in Table 2).

**MathQA on Wikidata (Comparison to Commercial Systems)** As for modes 3, 6, 9, and 12, we retrieved the formula results for mode 15 from Wikidata using a SPARQL query. Besides comparing it to the other index sources (arXiv and Wikipedia), we carried out an additional competition against a commercial knowledge base (Wolfram Alpha) and commercial search engine (Google). Mode 15 is the only one in which MathQA can be compared to its external competitors, as they do not allow for the other modes. Fig-ure 4 shows screenshots of the results of the different competing systems for an example formula (GoldID 363) from our test set. MathQA (above) and Google (below) display the same formula, but only MathQA allows for calculation. Wolfram Alpha (middle) also allows for calculation but using a different formula with different identifiers, which is relevant too. Table 4 lists the formulae retrieved by the three systems for the first five example query concept names (GoldID 310-315 in Table 1). For each system and GoldID, we evaluated whether a formula is displayed and relevant. Besides, the availability of a calculation is assessed. For Google, we additionally report the availability of a box around the formula in contrast to only highlighting the formula result in the text of a web page. The Google formula box is only available in the English language at the moment. The results of our investigation can be found in the evaluation folder at `https://github.com/ag-gipp/MathQA/blob/master/evaluation/semanticsearch/evalresultsMQAvsWAvsG.pdf`. For clarity, the coloring scheme is the same in Figure 4, Table 4, and the result table. Wolfram Alpha yields relevant formulae in 48% of the queries, 81% of which can be used to calculate the occurring quantities. MathQA performs slightly better with 52% relevant hits and almost equal 80% calculation availability. Google provides relevant results in 68% of the cases. However, only in 58%, a boxed formula is displayed and thus performing comparably to the other systems. Still, Google slightly outperforms the other two, which is not surprising given its expertise and budget. Yet, Google only in one case (GoldID 218: 'circumference'), i.e., 2% the possibility to calculate is enabled. One reason why MathQA could often not retrieve and display a formula is that for many of the cases, it links to Wikidata disambiguation page items. For example, 'work' can be either 'energy transferred to an object via the application of force on it through a displacement' (Q42213) or 'physical or virtual object made by humans' (Q386724). Both items have the same name.

Table 4: Query results for formula name questions. For the first five benchmark examples, the formula retrieved by MathQA (blue) is compared to the results of a commercial knowledge base (Wolfram Alpha, pink) and search engine (Google, lime).

| Query Concept Name | MathQA Formula | Wolfram Alpha Formula | Google Formula |
|---|---|---|---|
| acceleration | $a = dv/dt$ | $v = at$ | $\bar{a} = \Delta v / \Delta t$ |
| angular acceleration | $\alpha = d\omega/dt$ | $\omega = \alpha t$ | $\alpha = \Delta\omega/\Delta t$ |
| angular frequency | $\omega = 2\pi f$ | $\nu = \omega/(2\pi)$ | $\omega = 2\pi/t$ |
| angular momentum | $L = r \times p$ | $L = I\omega, \omega = 2\pi n$ | $L = mvr$ |
| angular velocity | $\omega = d\varphi/dt \cdot u$ | $\omega = \alpha t$ | $\omega = \Delta\theta/\Delta t$ |

Figure 4: Screenshots of MathQA (above), Wolfram Alpha (middle), and Google (below) answering the same question: "what is the formula for speed?". Coloring will be reused in Table 4.

**Comparison** Table 2 shows the results of modes 13-15, comparing top1 accuracy and mean DCG. Comparing the index sources, we find that, like for modes 7-12, Wikidata performs better than Wikipedia and the arXiv. This supports our assumption that formula name indices created from smaller corpora are more precise, in contrast to the results for the identifier name-symbol mappings. A question like 'What is the formula for symbol E?' would not be reasonable due to the large symbol ambiguity. This is why we did not evaluate the respective additional modes 16-18. As for the identifier relationship questions (modes 7-12), Wikidata performs best and is therefore also deployed for the formula concept name questions (modes 13-15) in the live version of MathQA.

## 6 Discussion

In this section, we describe our dataset benchmarking and discuss challenges with data and format impermanence of Wikidata.

### 6.1 Benchmarking

We introduce a benchmark for mathematical question answering in 15 evaluation modes on open formula data (Wikidata, Wikipedia, arXiv). The selection and results are persisted in the *MathMLben* and *MathQA* repositories (see Section 5). In the live version of our system, we deployed the best performing modes: Wikidata queries using Pywikibot (formula string retrieval via formula concept name) and Wikidata SPARQL queries (formula string retrieval via identifier names). After paper publication, we will add a reference to our sample dataset, evaluation metrics, and results to benchmark platforms, such as Papers With Code[23]. The repositories contain all necessary data persisted to reproduce the results and potentially compare and present improved systems. This is required since Wikipedia and Wikidata are constantly changing. To reproduce results on raw open data, the respective dumps of Wikipedia[24] and Wikidata [25] can be employed. For queries on data dumps, Tanon et al. introduced [42] a SPARQL endpoint for Wikidata history[26]. The impermanence of the data model and content change of Wikidata is discussed in the next section.

---
[23]https://paperswithcode.com/task/mathematical-question-answering
[24]https://archive.org/details/enwiki-20210120
[25]https://archive.org/details/wikibase-wikidatawiki-20210120

## 6.2 Challenges and Limitations

**Wikidata data model**  Employing open databases, such as Wikidata for information retrieval tasks and systems, such as question answering, has the advantage of profiting from a constantly growing community-curated collection of world knowledge. However, there are some drawbacks, such as data model impermanence and content change.

In the case of MathQA working on Wikidata, we noticed three challenges that prevented the system from constantly providing the same results:

- Wikidata users deleted the 'defining formula' for some items. For example, the equation $PV = nRT$ was shifted from the item 'gas' (Q11432) to 'ideal gas law' (Q191785), refining the semantic context. This forced us to change the list of formula name example questions.

- Wikidata users changed the formula identifier data model in both property usage (from 'has part' to 'in defining formula' / 'symbol represents') and sequence (from qualifier, item, symbol to qualifier, symbol, item). See Figure 5 for an illustration of the variants. This broke the functioning of the relationship questions until code adaption.

- In the geometry items, the object attributes can either be modeled as direct properties, e.g., 'area' (P2046) or 'volume' (P478) in the form property, formula. However, a commonly used alternative is linking the attribute items instead as 'has quality' (P1552) property, e.g., 'area' (Q11500) or 'volume' (Q39297) in the form property, item, formula. The emergence of further alternatives potentially breaks the geometry question functionality.

In summary, formulae can always be deleted from or shifted to other items. Furthermore, different properties may be used to store identifier semantics or geometric attributes. Aggravatingly, different hierarchical sequence schemes may be employed (see Figure 5). Notably, many attributes are available both as property and item (e.g., 'volume' as P1552 and Q39297). Changes in property usages and schemes are ideally discussed by the community on property talk [27] and proposal[28] pages.

To make our system more robust, we introduce a cache with an alert in case changes cause a previously working query to be broken. Users can then inspect the respective items to adapt the system. An interesting research question would be to track such changes to predict potential extensions using rule-based or statistical machine learning. This will help the system to auto-repair, searching for identifier information by employing semantic similarity metrics. Given data-query pairs, we can explore predicting one from the other.

---

[26] https://github.com/Tpt/wikidata-sparql-history  
[27] https://www.wikidata.org/wiki/Property_talk:P4934  
[28] https://www.wikidata.org/wiki/Wikidata:Property_proposal/symbol_represents

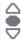

Figure 5: Variants in the Wikidata formula identifier data model (as of December, 29th 2021) in different schemes a) .qualifier, item, symbol, and b) qualifier, symbol, item.

**Further challenges** An additional challenge is the aforementioned issue that for some concept names, disambiguation page items (e.g., for 'work') prevent the system from finding the mathematical item. Furthermore, the Stanford parser sometimes does not provide the correct tree, especially for long concept names involving verbs or adjectives, such as 'Dirac equation in curved spacetime (Q16853908)'. Lastly, some concepts have synonyms that should link to the same formula concept, such as 'electric force' and 'Coulomb force' or 'M–sigma relation' and 'Faber-Jackson law.'

# 7 Conclusion and Outlook

In this section, we summarize our contributions and outline the benefits and future directions of our work.

## 7.1 Conclusion

In this paper, we demonstrated how Mathematical Entity Linking (MathEL) and Unsupervised Formula Labeling (UFL) can be used for semantic formula search and mathematical question answering. We implemented a system that can answer factoid natural language questions yielding a formula. The system also displays names of the constituting formula identifiers and values of constants if this information is available on Wikidata. Moreover, it allows for a computation using additional input values from the user. We tested our system and retrieval methods on a selection of annotated formula concepts, created from physics Wikipedia articles using a formula and identifier name annotation recommender system. We evaluated over 5,000 results in 15 different modes using identifier names and symbols as input or output, respectively, or formula names as input and strings as output, and Wikidata, Wikipedia, or the arXiv as source for the index.

**Results** Research questions 1-3 are answered in Table 2 and Section 5.2 to 5.4. In the identifier name and symbol index evaluations, Wikidata outperformed Wikipedia and the arXiv as source in terms of both top1 accuracy and Discounted Cumulative Gain (research question 4). Based on the results, we decided to choose Wikidata as source for the live version of MathQA since it is intended to only display the highest-ranked formula. Besides, in contrast to the other indices, which were fixed snapshots on the NTCIR 11/12 benchmark evaluation datasets, Wikidata is constantly extended by new formulae. For the search of formula strings by formula names, the Wikidata SPARQL query retrieval scored highest compared to the semantic indices created from the arXiv and Wikipedia. Using this best-performing mode, we compared our MathQA system to a commercial knowledge-base and calculation-engine (Wolfram Alpha) and search-engine (Google). While our system (52%) was outperformed by Google (68%), it outperformed Wolfram Alpha (48%). For 80% of the test formulae, MathQA could allow for calculations (research question 5). The advantage of our system over the commercial competitors is its transparency - being open source and working on linked open data. Moreover, Wolfram Alpha and Google can only answer formula concept name questions (modes 13-15) and not identifier queries (modes 1-12). MathQA is available hosted by Wikimedia at https://mathqa.wmflabs.org.

## 7.2 Outlook

Our MathQA system is intended to aid students and researchers from STEM disciplines in finding formulae by querying concept names or identifier relationships. To the best of our knowledge, there is no comparable search engine available so far. Students can get an overview of identifier relationships to understand connections between different identifier concepts better.

**Future Work** So far, we did not implement the question type 'What is the name of [formula string]?' This is the classical mode of a formula search engine, and we concentrate on the reverse modes, i.e., searching formula strings by names. Although this is not the focus of our research, we will extend this functionality. Moreover, we aim to automate the index construction from arXiv datasets or Wikipedia data dumps. MathQA should also return links to arXiv papers or Wikipedia articles (possibly with the surrounding text passage). However, as classical commercial search engines (e.g., Google) already provide this, it is again not our focus. What we consider a more important next step is to include the units of the identifiers [36], e.g., 'Coulomb' for 'charge.' However, the Wikidata knowledge-graph still needs to be completed with this information.

Having crawled formula and identifier relationships, we can create a graph of relevance relations, a 'FormulaRank' (in analogy to 'TextRank' [22] and 'PageRank' [25]) to get a concept map of a specific subject ontology (e.g., in physics). Lastly, Entity Linking to Wikidata items can possibly be used to support and enhance mathematical document classification [35] by augmenting subject class labels with concept labels. We will explore this in an upcoming research project.

# References

[1] A. Abdi, N. Idris, and Z. Ahmad. "QAPD: an ontology-based question answering system in the physics domain". In: *Soft Comput.* 22.1 (2018), pp. 213–230.

[2] A. Aizawa et al. "NTCIR-11 Math-2 Task Overview". In: *NTCIR*. National Institute of Informatics (NII), 2014.

[3] A. Amini et al. "MathQA: Towards Interpretable Math Word Problem Solving with Operation-Based Formalisms". In: *NAACL-HLT (1)*. Association for Computational Linguistics, 2019, pp. 2357–2367.

[4] H. Bast, B. Buchhold, and E. Haussmann. "Semantic Search on Text and Knowledge Bases". In: *Found. Trends Inf. Retr.* 10.2-3 (2016), pp. 119–271.

[5] A. Belz, S. Mille, and D. M. Howcroft. "Disentangling the Properties of Human Evaluation Methods: A Classification System to Support Comparability, Meta-Evaluation and Reproducibility Testing". In: *INLG*. Association for Computational Linguistics, 2020, pp. 183–194.

[6] J. Berant et al. "Semantic Parsing on Freebase from Question-Answer Pairs". In: *EMNLP*. ACL, 2013, pp. 1533–1544.

[7] A. Bhattacharya. "A Survey of Question Answering for Math and Science Problem". In: *CoRR* abs/1705.04530 (2017).

[8] K. D. Bollacker, R. P. Cook, and P. Tufts. "Freebase: A Shared Database of Structured General Human Knowledge". In: *AAAI*. AAAI Press, 2007, pp. 1962–1963.

[9] A. Bordes et al. "Large-scale Simple Question Answering with Memory Networks". In: *CoRR* abs/1506.02075 (2015).

[10] S. Cucerzan and E. Agichtein. "Factoid Question Answering over Unstructured and Structured Web Content". In: *TREC*. Vol. 500-266. National Institute of Standards and Technology (NIST), 2005.

[11] S. C. Deerwester et al. "Indexing by Latent Semantic Analysis". In: *J. Am. Soc. Inf. Sci.* 41.6 (1990), pp. 391–407.

[12] A. A. Gunawan, P. R. Mulyono, and W. Budiharto. "Indonesian question answering system for solving arithmetic word problems on intelligent humanoid robot". In: *Procedia Computer Science* 135 (2018), pp. 719–726.

[13] M. Hopkins et al. "SemEval-2019 Task 10: Math Question Answering". In: *SemEval@NAACL-HLT*. Association for Computational Linguistics, 2019, pp. 893–899.

[14] X. Huang et al. "Knowledge Graph Embedding Based Question Answering". In: *WSDM*. ACM, 2019, pp. 105–113.

[15] K. Järvelin and J. Kekäläinen. "Cumulated gain-based evaluation of IR techniques". In: *ACM Trans. Inf. Syst.* 20.4 (2002), pp. 422–446.

[16] H. Ji and R. Grishman. "Knowledge Base Population: Successful Approaches and Challenges". In: *ACL*. The Association for Computer Linguistics, 2011, pp. 1148–1158.

[17] G. Y. Kristianto and A. Aizawa. "Linking Mathematical Expressions to Wikipedia". In: *SWM@WSDM*. ACM, 2017, pp. 57–64.

[18] G. Y. Kristianto, G. Topic, and A. Aizawa. "Entity Linking for Mathematical Expressions in Scientific Documents". In: *ICADL*. Vol. 10075. Springer, 2016, pp. 144–149.

[19] C. Liang et al. "A Meaning-Based Statistical English Math Word Problem Solver". In: *NAACL-HLT*. Association for Computational Linguistics, 2018, pp. 652–662.

[20] B. Mansouri, R. Zanibbi, and D. W. Oard. "Characterizing Searches for Mathematical Concepts". In: *JCDL*. IEEE, 2019, pp. 57–66.

[21] B. Mansouri et al. "Finding Old Answers to New Math Questions: The ARQMath Lab at CLEF 2020". In: *ECIR (2)*. Vol. 12036. Springer, 2020, pp. 564–571.

[22] R. Mihalcea and P. Tarau. "TextRank: Bringing Order into Text". In: *EMNLP*. ACL, 2004, pp. 404–411.

[23] T. T. Nguyen, K. Chang, and S. C. Hui. "A math-aware search engine for math question answering system". In: *CIKM*. ACM, 2012, pp. 724–733.

[24] B. Ojokoh and E. Adebisi. "A Review of Question Answering Systems". In: *J. Web Eng.* 17.8 (2019), pp. 717–758.

[25] L. Page et al. *The PageRank citation ranking: Bringing order to the web*. Tech. rep. Stanford InfoLab, 1999.

[26] D. C. Pineau. "Math-Aware Search Engines: Physics Applications and Overview". In: *CoRR* abs/1609.03457 (2016).

[27] N. Prange. *WikiQuestions - A large question dataset generated from Wikipedia sentences*. Accessed: 2021-02-13.

[28] P. Rajpurkar et al. "SQuAD: 100, 000+ Questions for Machine Comprehension of Text". In: *EMNLP*. The Association for Computational Linguistics, 2016, pp. 2383–2392.

[29] P. Scharpf, M. Schubotz, and B. Gipp. "Fast Linking of Mathematical Wikidata Entities in Wikipedia Articles Using Annotation Recommendation". In: *Proceedings of the Web Conference (WWW) 2021*. ACM, Apr. 2021. DOI: 10.1145/3442442.3452348.

[30] P. Scharpf, M. Schubotz, and B. Gipp. "Mathematics in Wikidata". In: *Wikidata@ISWC*. Vol. 2982. CEUR-WS.org, 2021.

[32] P. Scharpf, M. Schubotz, and B. Gipp. "Representing Mathematical Formulae in Content MathML using Wikidata". In: *BIRNDL@SIGIR*. Vol. 2132. CEUR-WS.org, 2018, pp. 46–59.

[33] P. Scharpf et al. "*AnnoMath TeX* - a formula identifier annotation recommender system for STEM documents". In: *RecSys*. ACM, 2019, pp. 532–533.

[34] P. Scharpf et al. "ARQMath Lab: An Incubator for Semantic Formula Search in zbMATH Open?" In: *CLEF (Working Notes)*. Vol. 2696. CEUR-WS.org, 2020.

[35] P. Scharpf et al. "Classification and Clustering of arXiv Documents, Sections, and Abstracts, Comparing Encodings of Natural and Mathematical Language". In: *JCDL*. ACM, 2020, pp. 137–146.

[36] M. Schubotz, D. Veenhuis, and H. S. Cohl. "Getting the units right". In: *FM4M/MathUI/ThEdu/DP/WIP@CIKM*. Vol. 1785. CEUR-WS.org, 2016, pp. 146–156.

[37] M. Schubotz et al. "AutoMSC: Automatic Assignment of Mathematics Subject Classification Labels". In: *CICM*. Vol. 12236. Springer, 2020, pp. 237–250.

[38] M. Schubotz et al. "Improving the Representation and Conversion of Mathematical Formulae by Considering their Textual Context". In: *JCDL*. ACM, 2018, pp. 233–242.

[39] M. Schubotz et al. "Introducing MathQA: a Math-Aware question answering system". In: *Information Discovery and Delivery* 46.4 (2018), pp. 214–224.

[40] I. V. Serban et al. "Generating Factoid Questions With Recurrent Neural Networks: The 30M Factoid Question-Answer Corpus". In: *CoRR* abs/1603.06807 (2016).

[41] N. W. Smith. "A Question-Answering System for Elementary Mathematics." In: (1974).

[42] T. P. Tanon and F. M. Suchanek. "Querying the Edit History of Wikidata". In: *The Semantic Web: ESWC 2019 Satellite Events - ESWC 2019 Satellite Events, Portorož, Slovenia, June 2-6, 2019, Revised Selected Papers*. 2019, pp. 161–166. DOI: 10.1007/978-3-030-32327-1_32.

[43] T. P. Tanon et al. "Demoing Platypus - A Multilingual Question Answering Platform for Wikidata". In: *ESWC (Satellite Events)*. Vol. 11155. Springer, 2018, pp. 111–116.

[44] D. Vrandecic and M. Krötzsch. "Wikidata: a free collaborative knowledgebase". In: *Commun. ACM* 57.10 (2014), pp. 78–85.

[45] W. Yin et al. "Simple Question Answering by Attentive Convolutional Neural Network". In: *COLING*. ACL, 2016, pp. 1746–1756.

[46] R. Zanibbi et al. "Overview of ARQMath 2020: CLEF Lab on Answer Retrieval for Questions on Math". In: *CLEF*. Vol. 12260. Springer, 2020, pp. 169–193.

Listing 1: Use the following `BibTeX` code to cite this article

```
@inproceedings{Scharpf2022,
  author    = {Scharpf, Philipp and Schubotz, Moritz and
      Gipp, Bela},
  booktitle = {2022 ACM/IEEE Joint Conference on Digital
      Libraries (JCDL)},
  title     = {{M}ining {M}athematical {D}ocuments for {Q}
      uestion {A}nswering via {U}nsupervised {F}ormula {L}
      abeling},
  year      = {2022},
  month     = {June},
  location  = {Cologne, Germany},
  topic     = {mathir}
}
```